\documentclass[final,3p,times]{elsarticle}

 \usepackage{graphicx}

\usepackage{amssymb}
\usepackage{color}

\usepackage{siunitx}
\newcommand{\nuc}[2]{\hbox{$^{#1}$#2}}
\newcommand{\pips}{PIPS\textsuperscript{\textregistered}}
\newcommand{\comsol}{COMSOL Multiphysics\textsuperscript{\textregistered}}
\newcommand{\magnet}[3]{$#1^{\prime\prime}\times#2^{\prime\prime}\times#3^{\prime\prime}$}
\newcommand{\inch}{\ensuremath{^{\prime\prime}}}

\usepackage{gensymb}

\usepackage[hyphens]{url}
\usepackage{hyperref} 

\usepackage{tabularx}
\usepackage{booktabs}
\usepackage{float} 

\biboptions{sort,compress}
\journal{Nuclear Instruments and Methods in Physics Research, A}

\begin{document}

\begin{frontmatter}



\title{The ICESPICE demonstrator for particle/$\gamma$-$e^{-}$ coincidence experiments at Florida State University}

\author[fsu]{A.L.~Conley}
\address[fsu]{Department of Physics, Florida State University, Tallahassee, FL 32306, USA}

\author[fsu]{M.~Spieker\corref{cor1}}
\ead{mspieker@fsu.edu}
\cortext[cor1]{Corresponding author}

\author[fsu]{R.~Aggarwal}

\author[fsu]{L.T.~Baby}

\author[fsu]{J.~Davis}

\author[fsu]{J.~Esparza}

\author[fsu]{I.~Hay}

\author[fsu]{B.~Kelly}

\author[fsu]{T.~Kirk}

\author[fsu]{M.I.~Khawaja}

\author[lsu]{R.~Mahajan\fnref{label1}}
\fntext[label1]{Current address: Department of Physics and Astronomy, University of Kentucky, Lexington, Kentucky 40506-0055, USA}

\author[lsu]{S.T.~Marley}
\address[lsu]{Department of Physics and Astronomy, Louisiana State University, Baton Rouge, LA 70803, USA}

\author[fsu]{M.~Mestayer}

\author[fsu]{A.B.~Morelock}

\author[fsu]{A.~Peters}

\author[fsu]{A.M.~Ring}

\author[fsu]{J.~Sheridan}

\author[fsu]{V.~Sitaraman}

\author[fsu]{T.~Stuck}

\author[fsu]{I.~Wiedenh\"over}

\begin{abstract}
The Internal Conversion Electron SPectrometer In Coincidence Experiments (ICESPICE) demonstrator has been developed at Florida State University to enable particle/$\gamma$–$e^{-}$ coincidence measurements in low-energy nuclear structure studies. ICESPICE is based on the mini-orange spectrometer concept and features a modular design using commercially available permanent magnets arranged in toroidal configurations to transport internal conversion electrons to room-temperature \pips{} detectors while suppressing background from undesired particles. The system was optimized through SolidWorks modeling, \comsol{} magnetic field simulations, and Geant4 particle tracking to maximize the magnetic transmission probability for electrons around 1 MeV. Commissioning tests using a calibrated \nuc{207}{Bi} source demonstrated the performance of multiple spectrometer-detector configurations. Coincidence measurements between CeBr$_3$ detectors from the CeBrA array and \pips{} detectors revealed clear $\gamma$–$e^{-}$ correlations. The first in-beam particle–$e^{-}$ measurements using ICESPICE were performed with the Super-Enge Split-Pole Spectrograph (SE-SPS) in the \nuc{208}{Pb}$(d,t)$\nuc{207}{Pb} reaction. Prompt coincidences between tritons detected with the SE-SPS and electrons detected with ICESPICE were observed. The presented results show that ICESPICE is a promising ancillary detector system for in-beam internal conversion electron spectroscopy at the FSU SE-SPS.
\end{abstract}

\end{frontmatter}

\section{Introduction}

Internal conversion electron (ICE) spectroscopy is a powerful technique for investigating the structure of atomic nuclei. In this process, the nucleus de-excites by transferring its energy directly to an atomic electron, which is subsequently ejected. This internal conversion competes with $\gamma$ decay, and the relative probability of the two processes is quantified by the internal conversion coefficient ($\alpha$). The magnitude of $\alpha$ is sensitive to nuclear properties such as the atomic number, transition energy, and multipolarity of the transition between excited states involved. For low-energy or high-multipolarity transitions, especially in medium- and heavy-mass nuclei, internal conversion can dominate over $\gamma$ decay \cite{Krane:359790}. 

ICE spectroscopy plays a particularly crucial role in the study of electric monopole, $E0$, transitions between states of identical spin and parity, such as, {\it e.g.}, $0^+ \rightarrow 0^+$ \cite{kibedi_electric_2005} and $2^+ \rightarrow 2^+$ \cite{evitts_identification_2018} transitions. Measurements of $E0$ transition strengths provide direct insights into shape coexistence and configuration mixing, offering a sensitive probe for studying changes in nuclear deformation \cite{heyde_shape_2011, Gar22a}. Recent examples include the work of \cite{Se_1, Se_2, Se_3} in the Se isotopes.

Historically, a variety of spectrometer designs have been developed to enable in-beam ICE measurements. Early devices such as solenoidal and lens-type spectrometers offered good resolution but were limited by their large size, poor $\gamma$-ray suppression, and difficulty in implementing coincidence measurements \cite{witcher_electron_1941, mladjenovic_recent_1960}. The mini-orange spectrometer (MOS), created by van Klinken \cite{van_klinken_conversion_1972}, addressed these challenges by using permanent magnets arranged around a central absorber to produce a toroidal magnetic field. This compact, cost-effective design enables efficient transport of electrons across a selectable range of energies via different magnet configurations while suppressing backgrounds from $\gamma$ rays, $\delta$ rays, positrons, and heavier particles emitted from the target position. Its small footprint and effective background rejection make it especially suitable for in-beam $\gamma$–$e^-$ coincidence experiments. Modern MOS implementations include fIREBall \cite{lee_transformation_2023}, SLICES \cite{marchini_slices_2021}, and SPICE \cite{ketelhut_SPICE_2014}.

\begin{figure}[t]
    \centering
    \includegraphics[width=0.6\linewidth]{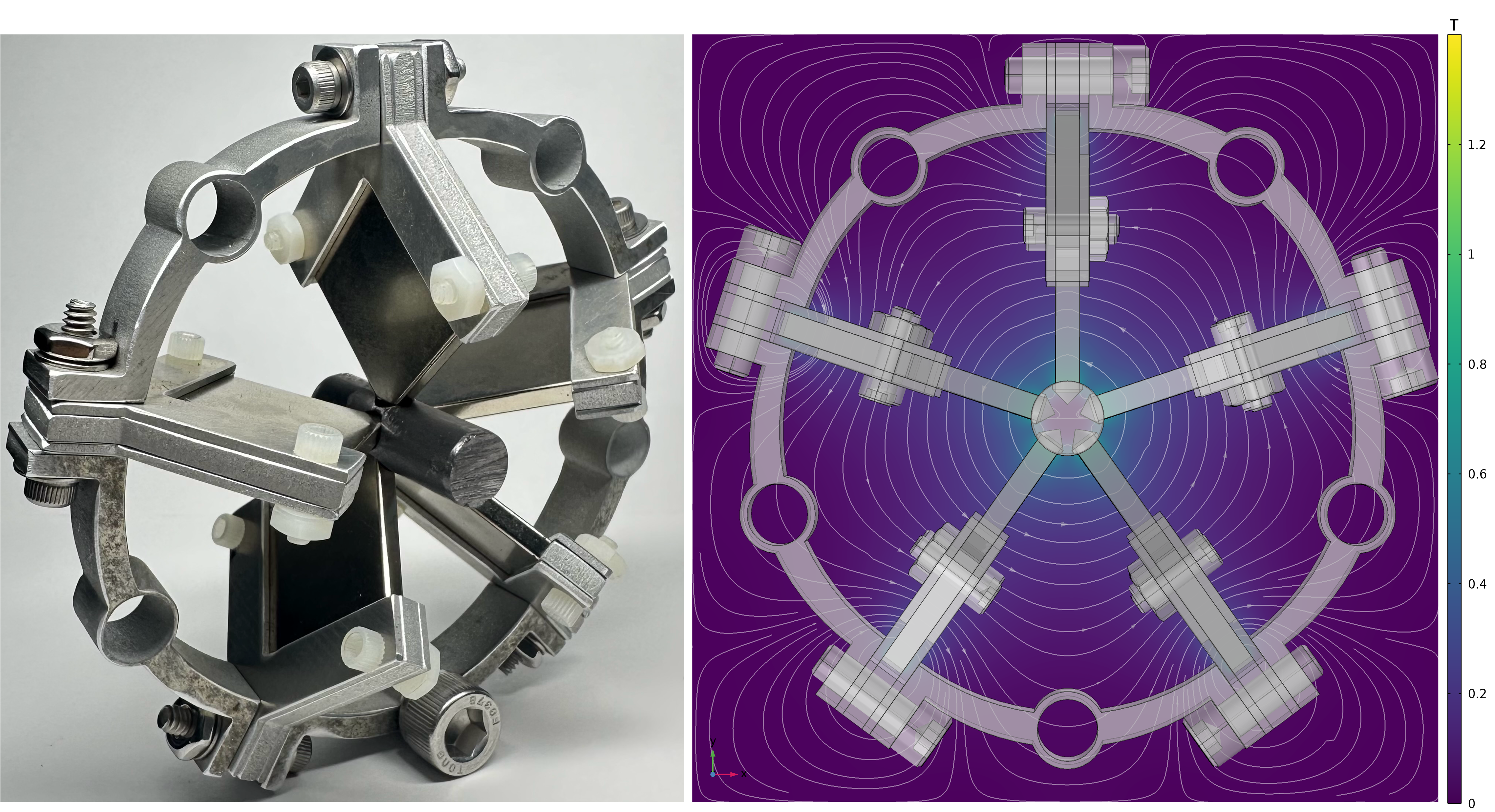}
    \caption{(left) Photograph of the first ICESPICE spectrometer configuration, consisting of five \magnet{1}{1}{1/8} N42 magnets arranged around a tantalum attenuator with a diameter of 3/8\inch and a length of 1.2\inch. (right) Two-dimensional \comsol{} simulation of the magnetic field for the same configuration. The region of greatest field homogeneity around the attenuator extends to a radius of approximately 18~mm.}
    \label{fig: comsol simulation + icespice photograph}
\end{figure}

ICESPICE (Internal Conversion Electron SPectrometer In Coincidence Experiments) builds on the mini-orange concept with a modernized yet simple and modular design tailored to in-beam ICE spectroscopy at the Super-Enge Split-Pole Spectrograph (SE-SPS) at Florida State University \cite{spieker_nuclear_2024}. When operated in coincidence with the Cerium Bromide Array (CeBrA) \cite{conley_cebra_2024}, ICESPICE enables $\gamma$–$e^-$ coincidence measurements. The spectrometer array was optimized using simulations in SolidWorks \cite{solidworks}, \comsol{} \cite{comsol}, and Geant4 \cite{geant4_cite1, geant4_cite2, geant4_cite3}. Each spectrometer incorporates a room-temperature \pips{} (Passivated Implanted Planar Silicon) detector \cite{noauthor_pips_nodate} placed behind a mini-orange magnetic filter. These detectors were successfully used for the detection of conversion electrons in \cite{ahmad_room-temperature_2015, ahmad_high_resolution_2015}. The toroidal magnetic fields generated by the permanent magnets selectively guide conversion electrons to the \pips{} detector, significantly reducing background, and enables efficient measurements across a tunable energy range using different magnet configurations. 


\begin{figure}[t]
    \centering
    \includegraphics[width=0.6\linewidth]{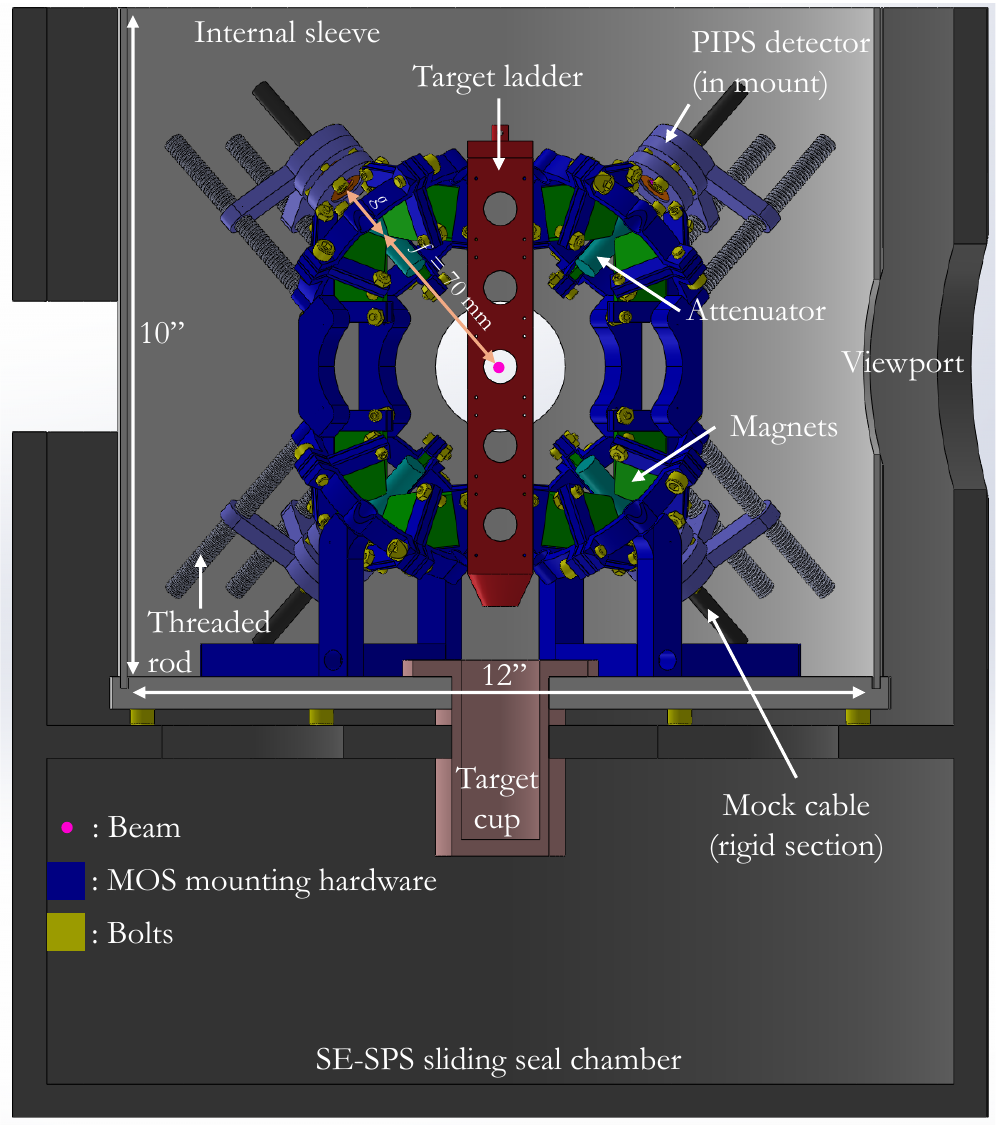}
    \caption{This sliced view of the SE-SPS sliding-seal chamber and internal sleeve shows the backward-angle geometry. The opening behind the target ladder corresponds to the incoming beam direction. Four mini-orange spectrometers (MOS) are shown positioned at backward angles within the 12-inch-diameter, 10-inch-tall sleeve, setting the minimum feasible target-to-spectrometer distance of $f = 70$ mm at $\phi = \pm45^\circ, \pm135^\circ$ and $\theta = \pm120^\circ$ without interfering with the viewport or beamline.}
    \label{fig:solidworks assmebly}
\end{figure}

\section{Optimization of ICESPICE}

The goal of optimizing ICESPICE was to maximize the magnetic transmission probability, defined as 

\begin{equation} \label{eqn: magnetic_transmission}
    T_{M}(E) = \frac{N^{\mathrm{ICESPICE}}_{e^-}(E)}{N^{4\pi}_{e^-}(E)},
\end{equation}

\noindent where $T_{M}(E)$ represents the transmission probability for electrons of energy $E$, $N^{\mathrm{ICESPICE}}_{e^-}$ is the number of electrons of energy $E$ that pass through the magnetic field and are detected within the defined active area of the detector, and $N^{4\pi}_{e^-}$ is the total number of electrons of energy $E$ emitted isotropically from the target position. Maximizing $T_{M}(E)$ involves optimizing the design of the magnetic filter to ensure that the maximum number of electrons emitted from the target position are transported to the detector. This process is constrained by the experimental setup and its physical limitations. 

\begin{figure}[t]
    \centering
    \includegraphics[width=0.6\linewidth]{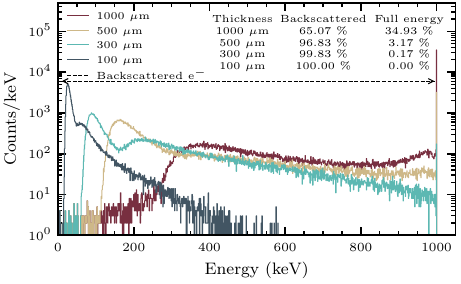}
    \caption{Simulated energy deposition spectra in PIPS\textsuperscript{\textregistered} detectors with varying thicknesses (100, 300, 500, and 1000~$\mu$m) for 10$^5$ monoenergetic 1000~keV electrons with an incident angle of $\theta=0^{\degree}$. The percentage of electrons depositing their full energy decreases significantly with thinner detectors, increasing the contribution of low-energy tails, i.e., of a continuum caused by backscattering of electrons that leave the active detector region.}
    \label{fig: PIPS backscattering}
\end{figure}

The initial design of ICESPICE was intentionally kept simple, inspired by the mini-orange spectrometers used in ICEBALL \cite{metlay_iceball_1993}. These spectrometers utilized six \magnet{1}{1}{1/8} SmCo$_5$ magnets arranged around a lead absorber \cite{metlay_iceball_1993}. Recently, Lee \textit{et al.} transformed ICEBALL into fIREBALL by enhancing the efficiency of the magnetic filters through the use of custom-shaped magnets from Electron Energy Corporation, supported by simulations \cite{lee_transformation_2023, electron_energy_corporation}. Similarly, mini-orange spectrometers such as \cite{zheng_CIAE_2016, marchini_slices_2021} have employed wedge-shaped magnets to achieve a more uniform toroidal magnetic field. While similar advanced magnet designs could be explored in the future, the ICESPICE demonstrator prioritizes cost-effective designs using readily available commercial magnets \cite{magnet_vendor}.

To determine the optimal magnet size and strength, extensive simulations were conducted. Different designs were created in SolidWorks \cite{solidworks} and imported into \comsol{} \cite{comsol} to simulate the magnetic field of different shapes and strengths of readily available magnets using the AC/DC module. The magnetic field was then exported as a 3-D map with a resolution of 0.5 mm. A 2-D snapshot of a design with five \magnet{1}{1}{1/8} N42 magnets around a tantalum attenuator with a diameter of 3/8\inch{} is shown in Figure~\ref{fig: comsol simulation + icespice photograph}. The exported 3D magnetic field map was imported into Geant4 for particle tracking and detector response simulations. While particle tracking can be performed in \comsol{} using the Particle Tracing Module, it is more computationally expensive than in Geant4. 
The SolidWorks geometry was easily imported into Geant4 using CADMesh \cite{cadmesh_geant4_poole2012acad}. The Geant4 simulation can be found at \cite{ICESPICE_github}.

All designs are subject to additional constraints that affect optimization of the transmission probability in a specific energy region. One key constraint is the geometry of the SE-SPS sliding seal chamber. An internal sleeve was recently constructed for this chamber to contain potential contamination of the scattering chamber during experiments with actinide targets. The spectrometers must, therefore, fit at backward angles within a sleeve of $\sim$12\inch{} diameter and 10\inch{} height, without obstructing the beam, target ladder, or the view port for beam focusing purposes. This limits the minimum target-to-spectrometer distance to $f=70$ mm, where $f$ is measured from the target to the spectrometer origin at the magnetic-field center (approximately at the magnet contact point with the Ta attenuator) [see Fig.\,\ref{fig:solidworks assmebly}], when positioned at azimuthal angles $\phi = \pm45^\circ, \pm135^\circ$ and polar angles $\theta = \pm120^\circ$, where $\theta = 0^\circ$ corresponds to the beam direction. The SolidWorks \cite{solidworks} assembly with four mini-orange spectrometers placed at backward angles, and including the internal sleeve and the SE-SPS sliding seal chamber is shown in Figure \ref{fig:solidworks assmebly}.

Furthermore, the detector geometry imposes more constraints. The distance spectrometer-to-detector (see Fig.\,\ref{fig:solidworks assmebly}), $g$, is restricted to 20--50~mm, and the geometric efficiency depends on the detector size. The available Passivated Implanted Planar Silicon (\pips{}) detectors at FSU have thicknesses of 100, 300, 500, and 1000~$\mu$m, all with an active area of 50~mm$^2$. The active area impacts the detection solid angle and, consequently, the magnetic transmission efficiency. The detector thickness significantly influences the ratio of the total absorption peak (full-energy deposition) to the low-energy continuum resulting from electron backscattering. Thus, thicker detectors are preferred for high-energy electron measurements due to their higher stopping power for electrons. Earlier, Ahmad \textit{et al.}~\cite{ahmad_room-temperature_2015, ahmad_high_resolution_2015} demonstrated that room-temperature \pips{} detectors of 100-, 300-, and 500-$\mu$m thickness and 25~mm$^2$ active area are well-suited for detecting low-energy internal conversion electrons from actinide nuclei. Their detectors achieved an energy resolution of $\sim$2.2~keV (FWHM) at 100~keV, comparable to cooled Si(Li) or PIN diodes, while exhibiting reduced $\gamma$- and X-ray sensitivity that yielded notably cleaner low-energy electron spectra. In their studies, Ahmad {\it et al.} measured electron energies of up to 400\,keV. In our work, we were interested to further test these \pips{} detectors for conversion electron detection and to see whether they could potentially be used for measuring 1-MeV electrons relevant for the study of $E0$ transitions between $J^{\pi} = 0^+$ states in the actinides \cite{Hoo96a, Spi13a}. To get an initial idea of the feasibility, $10^5$ monoenergetic electrons (1000~keV) were simulated with Geant4 for each of the available detector geometries. As shown in Figure~\ref{fig: PIPS backscattering}, only $\sim$35\% of electrons deposited their full energy in the 1000-$\mu$m detector. This fraction decreased even further to $\sim$3\% for the 500-$\mu$m detector and $\sim$0\% for the 300- and 100-$\mu$m detectors.

\begin{figure}[t]
    \centering
    \includegraphics[width=0.6\linewidth]{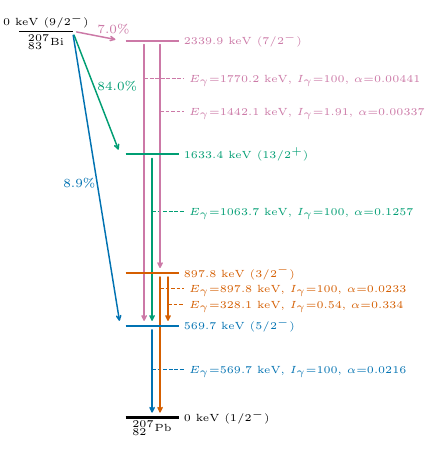}
    \caption{Decay scheme of \nuc{207}{Bi} to excited states in \nuc{207}{Pb} via electron capture (angled solid arrows), followed by $\gamma$-ray emission (vertical solid arrows). Levels are labeled with excitation energy ($E_x$) and $J^\pi$. Transition labels indicate the $\gamma$-ray energy $E_\gamma$, relative intensity $I_\gamma$, and internal conversion coefficient $\alpha$. The corresponding electron energies are listed in Table~\ref{tab:207Bi_decay}.}
    \label{fig:207Bi_decay}
\end{figure}

Based on the combined SolidWorks, \comsol{}, and Geant4 studies, the highest-performing configuration for $\sim$1 MeV electrons, within the SE-SPS geometric constraints, with readily available magnets, and using the currently available \pips{} detectors at FSU, consisted of five N42 \magnet{1}{1}{1/8} permanent magnets arranged around a 3/8\inch{} diameter tantalum attenuator, positioned at $f = 70$ mm and $g = 25$  to $35$ mm, and used the 1000-$\mu$m thick \pips{} detector. We decided to test the same magnet configuration with the 500-$\mu$m detector, which was expected to retain some total-absorption efficiency (see Fig.\,\ref{fig: PIPS backscattering}). The 300-$\mu$m detector, by contrast, was coupled to a mini-orange spectrometer optimized for lower-energy electrons, using five N42 \magnet{1}{1}{1/16} magnets and the same attenuator geometry. The measured transmission probabilities for these configurations are presented in the following section, in direct comparison to experimental results from a \nuc{207}{Bi} calibration source measurement.

\section{Source Testing with \nuc{207}{Bi}} \label{sec:207Bi Source Testing}

\begin{table}[t]
    \centering
    \scriptsize
    \caption{Selected $\gamma$-ray and internal conversion electron energies for the decay of \nuc{207}{Bi} to excited states in \nuc{207}{Pb} (see decay scheme in Figure~\ref{fig:207Bi_decay})~\cite{kondev_nuclear_2011} used to determine the transmission probabilities. The \nuc{207}{Bi} source was calibrated on February 4th, 2025 with an activity of 23(1)~kBq.}
    \label{tab:207Bi_decay}
    \begin{tabularx}{\linewidth}{l l X X}
        \toprule
        Source & $E_{\gamma}$ [keV] & $E_{CE}$ [keV] & Intensity [\%] \\
        \midrule
        \nuc{207}{Bi} & 569.698(2) & K: 481.6935(21) & 1.537(22) \\
                      &            & L: 553.8372(21) & 0.442(6)  \\
                      &            & M: 565.8473(21) & 0.111(5)  \\
                      & 1063.656(3)& K: 975.651(3)   & 7.08(17)  \\
                      &            & L: 1047.795(3)  & 1.84(5)   \\
                      &            & M: 1059.805(3)  & 0.44(3)   \\
                      & 1770.228(9)& K: 1682.224(9) & 0.0238(12) \\
        \bottomrule
    \end{tabularx}
\end{table}

Source testing used FSU’s open \nuc{207}{Bi} source, calibrated on February 4th, 2025 with an activity of 23(1) kBq. Each \pips{} detector was tested with both a CANBERRA 2003BT Silicon Detector Preamplifier \cite{mirion_2003bt_preamp} and a mesytec MPR-1 preamplifier \cite{mesytec_mpr1}. Initial tests, including $\gamma$–$e^{-}$ coincidence measurements (see Sec. \ref{sec: gamma-electron coincidence}), were conducted in a small vacuum chamber at $10^{-2}$ Torr. Subsequent source tests and preparation for the first in-beam particle–$e^{-}$ coincidence experiment, which is discussed in Sec. \ref{sec: ICESPICE+SE-SPS}, took place in the SE-SPS scattering chamber at $10^{-5}$ Torr.

\begin{figure*}[t]
    \centering
    \includegraphics[width=0.99\linewidth]{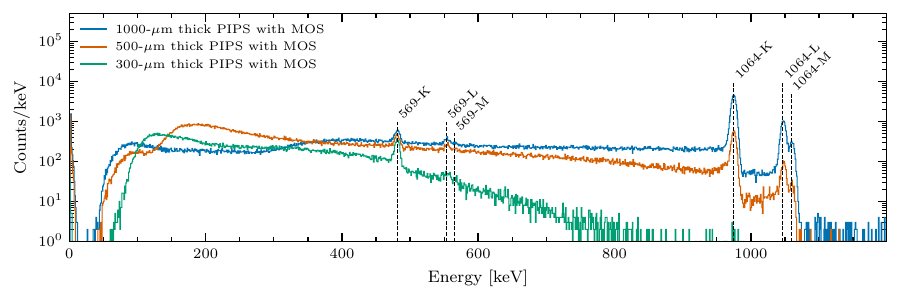}
    \caption{Spectra from the decay of \nuc{207}{Bi} measured using three PIPS\textsuperscript{\textregistered} detectors with different mini-orange configurations. All detectors were tested in identical geometry ($f = 70$ mm, $g = 25.4$ mm). The 300-$\mu$m detector was paired with a mini-orange spectrometer using five N42 \magnet{1}{1}{1/16} magnets, and the 500-$\mu$m and 1000-$\mu$m detectors with mini-orange spectrometers using five N42 \magnet{1}{1}{1/8} magnets. The achieved energy resolution (FWHM) was 8–10~keV for the electrons detected from the decay of the 1633-keV state in \nuc{207}{Pb}. The 20-hour measurement served as the calibration run for the first in-beam particle–$e^-$ coincidence tests (see Fig. \ref{fig:ICESPICE in SE-SPS Chamber} and Sec. \ref{sec: ICESPICE+SE-SPS}).}
    \label{fig:207Bi Spectrum with different PIPS}
\end{figure*}

\begin{figure}[t]
    \centering
    \includegraphics[width=0.6\linewidth]{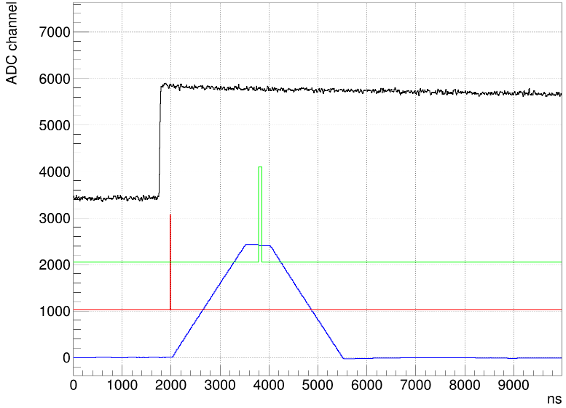}
    \caption{Digital filter windows for the 1000-$\mu$m thick PIPS\textsuperscript{\textregistered} detector coupled to a mesytec MPR-1 preamplifier~\cite{mesytec_mpr1} and digitized with a CAEN V1725S digitizer using proprietary DPP-PHA firmware \cite{caen}. The red trace represents the trigger, the blue trace shows the trapezoidal filter, and the green trace marks the sampling region at the peaking time.}
    \label{fig:pips1000_signal_mesytec_preamp}
\end{figure}

Exemplary electron spectra measured after electron capture of \nuc{207}{Bi} (see Figure~\ref{fig:207Bi_decay} for decay scheme) with the different \pips{} detectors are shown in Figure~\ref{fig:207Bi Spectrum with different PIPS}. These 20-hour long measurements were conducted in the SE-SPS scattering chamber and served as a calibration run for the first in-beam particle–$e^{-}$ coincidence experiment (see Sec. \ref{sec: ICESPICE+SE-SPS} for more details). All detectors were tested in identical geometry. The 300-$\mu$m thick \pips{} detector was paired with a mini-orange spectrometer using five N42 \magnet{1}{1}{1/16} magnets, and the 500-$\mu$m and 1000-$\mu$m \pips{} detectors with mini-orange spectrometers using five N42 \magnet{1}{1}{1/8} magnets. The energy resolution (FWHM), achieved in this work, was 8–10~keV for the conversion electron transitions coming from the 1633-keV state in \nuc{207}{Pb}. The 500-$\mu$m and 1000-$\mu$m thick \pips{} detectors used the mesytec MPR-1 preamplifier \cite{mesytec_mpr1} while the 300-$\mu$m thick detector used the CANBERRA 2003BT Silicon Detector Preamplifier \cite{mirion_2003bt_preamp}. The digitized signal for the 1000-$\mu$m thick \pips{} detector connected to a mesytec MPR-1 preamplifier \cite{mesytec_mpr1} and obtained with a CAEN V1725S digitizer with proprietary DPP-PHA firmware \cite{caen} is shown in Figure~\ref{fig:pips1000_signal_mesytec_preamp}. The red line indicates the trigger, while the blue line shows the trapezoidal filter output. The trapezoidal rise time was set to 1.488~$\mu$s and the flat-top duration to 0.496~$\mu$s. The sampling region, shown in green, averaged four samples at 50\% of the flat-top amplitude. A clear benefit of the 1000-$\mu$m over the 500-$\mu$m \pips{} detector is evident in Figure~\ref{fig:207Bi Spectrum with different PIPS}: under identical MOS settings and the same run time, the 500-$\mu$m detector yielded only 11\% of the 1064-keV K-conversion electrons compared to those detected with the 1000-$\mu$m \pips{} detector.

Realistic Geant4 simulations were carried out to benchmark the detector response against the experimentally measured \nuc{207}{Bi} spectra. The simulations included a detailed description of the experimental geometry: the source backing, testing chamber, all mounting hardware for the MOS, and the \pips{} detector window were modeled based on manufacturer data sheets \cite{noauthor_pips_nodate}. A production cut of 1~$\mu$m was applied throughout, and the \texttt{G4EmStandardPhysics\_option4} physics list was used. The source itself was implemented as a cylinder with a diameter of 5~mm and thickness of 1000~nm. The General Particle Source (GPS) was employed together with the \texttt{G4GeneralIon} to define the \nuc{207}{Bi} ground-state ion, with the subsequent decay handled by the \texttt{G4RadioactiveDecay} process using the internal decay data provided within Geant4.

\begin{figure}[t]
    \centering
    \includegraphics[width=0.5\linewidth]{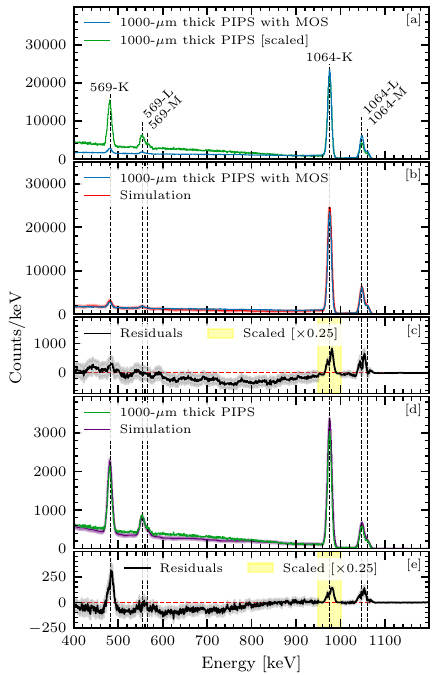}
    \caption{a) Measured \nuc{207}{Bi} spectrum with and without a the mini-orange spectrometer (MOS) configuration with five N42 \magnet{1}{1}{1/8} for the 1000-$\mu$m thick PIPS\textsuperscript{\textregistered} detector with a 50~mm$^2$ active area in identical geometry ($f = 70$ mm, $g = 30$ mm). The spectrum without the MOS was normalized to the yield of the 1064-keV K-shell conversion electrons. Panels (b) and (d) compare the detector response with and without the MOS, respectively, to the Geant4 simulation, while panels (c) and (e) show the residuals between simulation and experiment. The gray band shows the 1$\sigma$ statistical uncertainty on the residuals, combining the simulation and experimental uncertainties.}
    \label{fig:geant4 sim vs exp}
\end{figure}

Figure~\ref{fig:geant4 sim vs exp} compares the measured and simulated electron spectra after electron capture of \nuc{207}{Bi} for the 1000-$\mu$m thick detector. Panel (a) shows the measured spectra with and without the MOS. Each measurement was a 46-hour run, and the spectrum without the MOS was normalized to the 1064-keV K-conversion line. As expected, the MOS focuses electrons with energies of around 1\,MeV and suppresses conversion electrons with lower energies. Panels (b) and (d) compare the measured detector response with Geant4 simulations for the configurations with and without the MOS, respectively, while panels (c) and (e) show the corresponding residuals. The simulated spectra were scaled to minimize the residuals between 400 and 1200~keV. The residuals show that the simulation overpredicts the full-energy deposition for the 1064-keV K, L, and M electrons. This suggests that the probability for electrons to deposit their full energy is slightly lower in experiment than in the simulation. Consequently, experimental losses due to backscattering are higher, which is consistent with the residual structure showing an excess of low-energy counts in the data relative to the simulation.

\begin{figure}[t]
    \centering
    \includegraphics[width=0.6\linewidth]{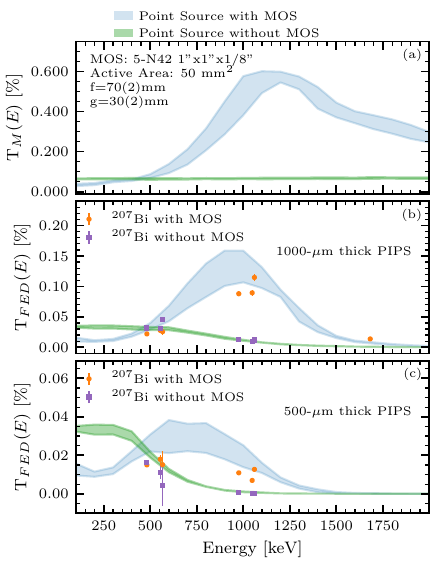}
    \caption{(a) Magnetic transmission probability $T_{M}(E)$ from Geant4 simulations for a detector with an active area of 50 mm$^2$ positioned at $f = 70(2)$ mm and $g = 30(2)$ mm relative to the source, with and without a mini-orange spectrometer (MOS) consisting of five N42 \magnet{1}{1}{1/8} magnets. The shaded band reflects uncertainties from varying the detector positions by $\pm 2$\,mm in the simulation. A point source was simulated with $10^6$ electrons emitted isotropically in 100 keV steps. (b) Full-energy deposition transmission probability $T_{\mathrm{FED}}(E)$ for a 1000-$\mu$m thick \pips{} detector (PD50-12-1000AM). Experimental data from the 46-hour \nuc{207}{Bi} measurement (see Fig.\ref{fig:geant4 sim vs exp}) are compared with simulations for both MOS and non-MOS configurations. (c) Same as panel (b), but for a 500-$\mu$m thick detector (PD50-11-500AM).}
    \label{fig:pips_trans_prob}
\end{figure}

Figure~\ref{fig:pips_trans_prob} presents the transmission probabilities determined from the Geant4 simulations and the experimental data. Panel (a) shows the magnetic transmission probability $T_{M}(E)$ for a 50-mm$^2$ active area; the MOS enhances transmission around 1~MeV, with the uncertainty band reflecting variations in the detector distances $f$ and $g$. Due to the current mounting scheme, $f$ and $g$ are known only to $\pm$2~mm. Because of the active area of the \pips{} detectors being small, tight focusing is required to optimize transmission, making $T_{M}(E)$ highly sensitive to these distances. Panels (b) and (c) show the full-energy deposition transmission probability $T_{\mathrm{FED}}(E)$ for the 1000-$\mu$m and 500-$\mu$m detectors, respectively. While the simulations reproduce the overall spectral trends, discrepancies remain, likely arising from differences between the modeled magnetic field (from COMSOL) and the true field. Future improvements may be achieved by directly mapping the magnetic field. Overall, the simulations reproduce the essential transmission behavior of ICESPICE.

\begin{figure*}[t]
    \centering
    \includegraphics[width=0.8\linewidth]{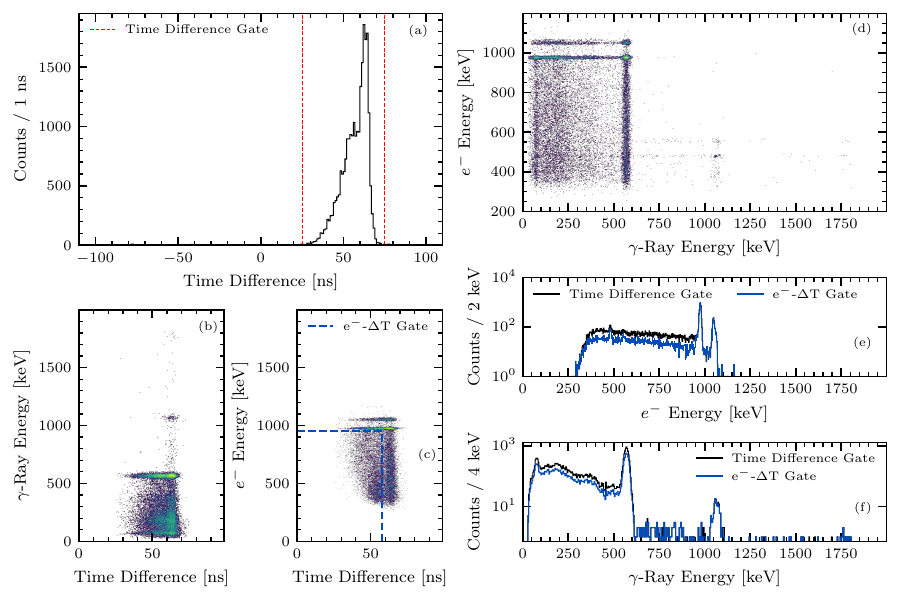}
    \caption{
    (a) Time-difference, $\Delta t$, spectrum between a 3\inch{}~$\times$~4\inch{} CeBr$_3$ detector and a 1000~$\mu$m-thick \pips{} detector coupled to a mini-orange spectrometer (MOS) configured with five N42 \magnet{1}{1}{1/8} magnets. Time differences are defined relative to the CeBr$_3$ detector. The measurement lasted 46~hours with ICESPICE positioned at $f = 70(2)$~mm and $g = 30(2)$~mm. The CeBr$_3$ detector was located at $\theta =90^\circ$ and at a distance of approximately 4~inches from the \nuc{207}{Bi} source. A clear prompt coincidence peak is visible. Panel (b) [c] shows the CeBr$_3$  $\gamma$-ray [\pips{} electron] energy vs the time difference, revealing the prompt and delayed structures. The $e^{-}-{\Delta}T$ gate selects data outside of the region marked by the the dashed line. This gate rejects $e^-$ events with partial energy deposition. Panel (d) shows the time-difference gated $\gamma$--$e^{-}$ coincidence matrix for the decay of \nuc{207}{Bi}. Multiple decay paths can be seen. For example, the 570-keV $\gamma$ ray is detected in coincidence with electrons originating from the decay of the 1633-keV, $J^{\pi} = 13/2^+$ state in \nuc{207}{Pb} (see Fig.\,\ref{fig:207Bi_decay}). Projections of this matrix onto the $e^-$ and $\gamma$-ray energy axes are shown in panel (e) and (f). These panels also show how the $e^{-}-{\Delta}T$ gate can provide additional background reduction.
    }
    \label{fig:207Bi electron-gamma coincidence matrix}
\end{figure*}

\subsection{$\gamma$-$e^{-}$ Coincidences using CeBrA} \label{sec: gamma-electron coincidence}

\begin{figure*}[t]
    \centering
    \includegraphics[width=.8\linewidth]{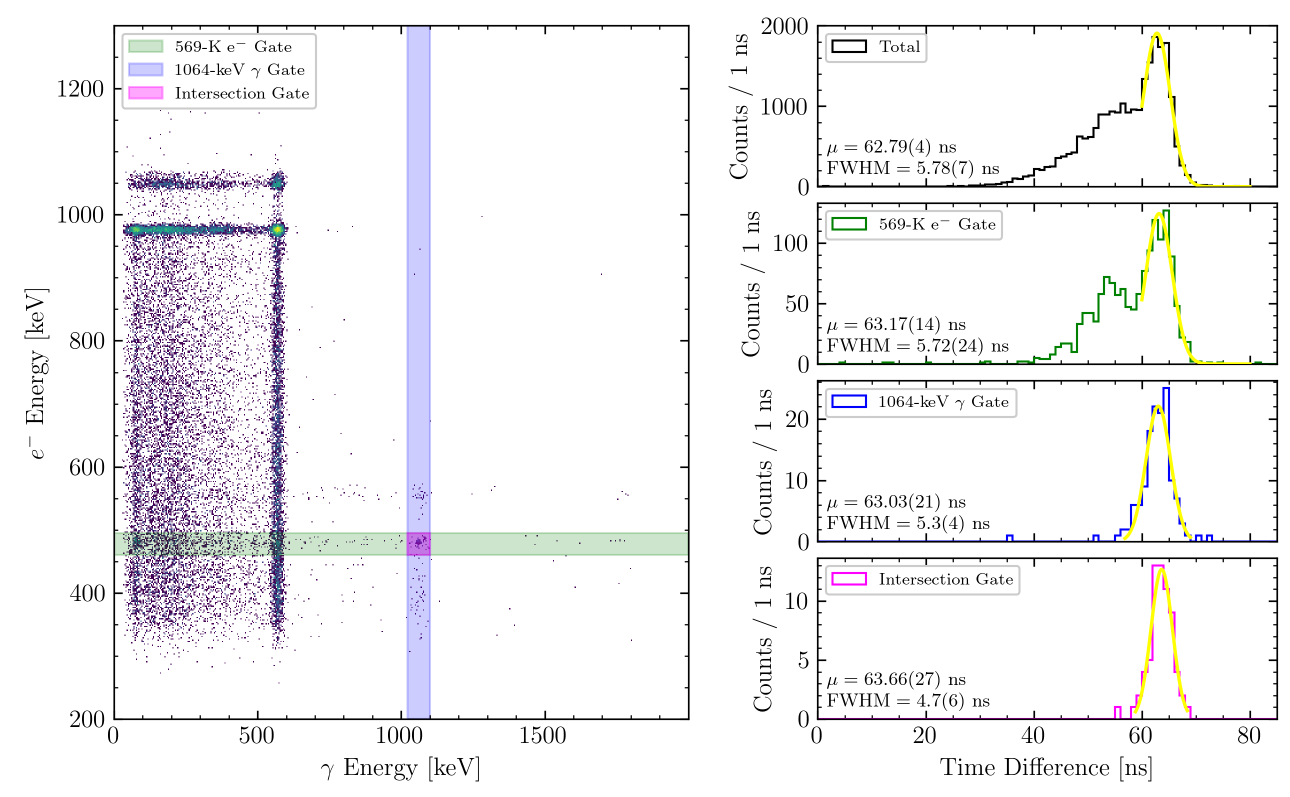}
    \caption{Electron-$\gamma$ coincidence matrix (left), also shown in Fig.\,\ref{fig:207Bi electron-gamma coincidence matrix}\,(d), and time-difference spectra between 1000-$\mu$m PIPS and $3 \times 4$ inch CeBr$_3$ detector (right) when energy gates indicated in the panels are applied. Gaussian fits to the ``prompt'' part of the timing distribution are shown. The mean position, $\mu$, of the Gaussian does not change for the different gates. The 1064-keV $\gamma$-ray gate and ``intersection'' gate establish the prompt coincidence timing resolution of $\Delta t = 5.0(5)$\,ns between the 1000-$\mu$m PIPS and $3 \times 4$ inch CeBr$_3$ detector. The resolution is consistent with those obtained with the other gates, which, however, contain contributions from electron scattering events in the PIPS with partial energy deposition leading to a tail in the timing distribution.}
    \label{fig:207Bi_new_timing}
\end{figure*}

The first $\gamma$-$e^{-}$ coincidence measurement with ICESPICE used a 3$''\times$4$''$ CeBr$_3$ $\gamma$-ray detector from the CeBrA demonstrator \cite{conley_cebra_2024}, a 1000-$\mu$m \pips{} detector, and a mini-orange spectrometer with five N42 \magnet{1}{1}{1/8} magnets using $f = 70$ mm and $g = 30$ mm. Clear coincidences between $\gamma$ rays and conversion electrons emitted after the electron-capture decay of \nuc{207}{Bi} were observed, see Figure~\ref{fig:207Bi electron-gamma coincidence matrix}. 

\begin{figure}[t]
    \centering
    \includegraphics[width=0.3\linewidth]{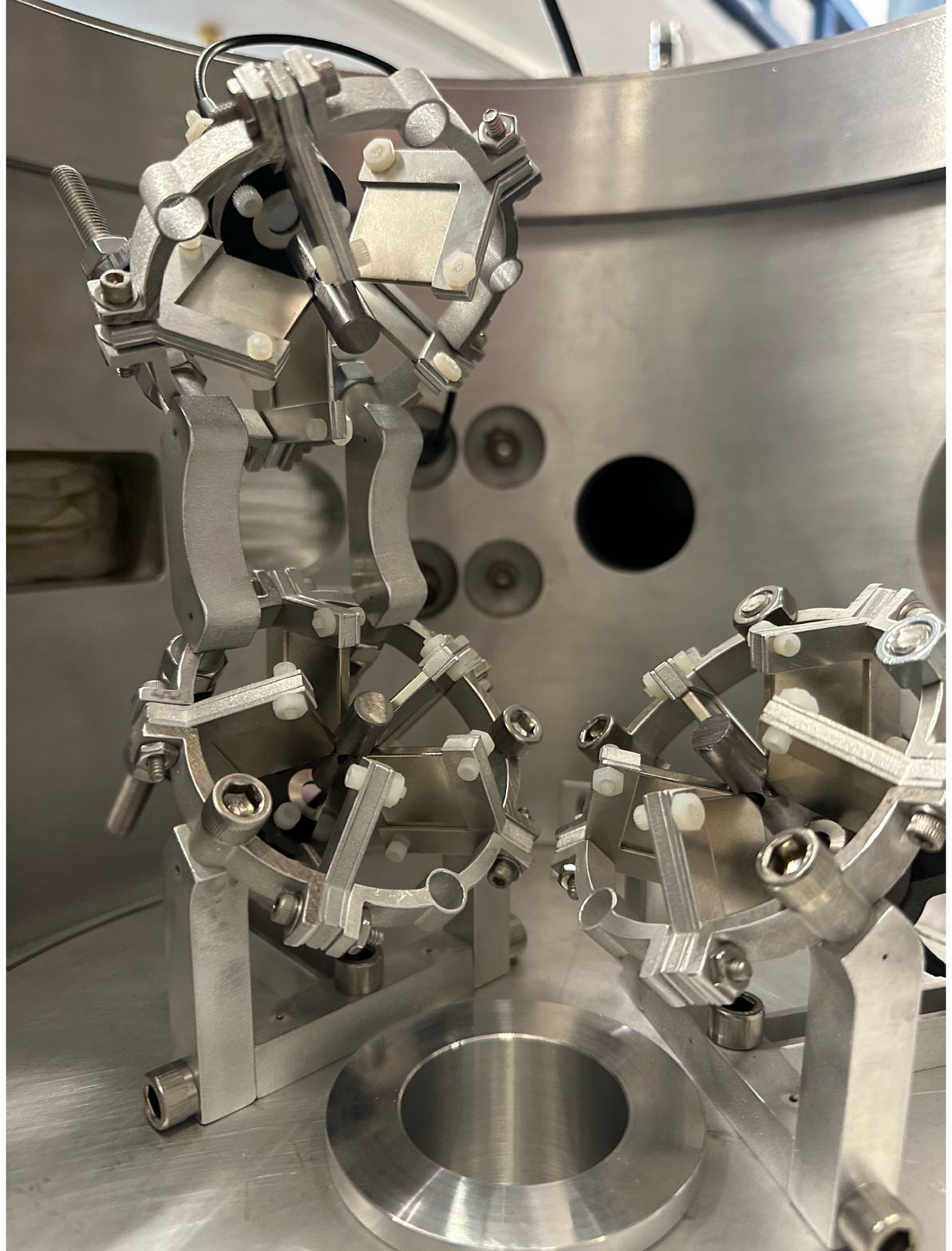}
    \caption{Photograph of the first in-beam particle-$e^-$ coincidence test using ICESPICE and the SE-SPS. Three mini-orange spectrometers were installed at backward angles. One spectrometer (top-left) used five N42 \magnet{1}{1}{1/16} magnets with a 300-$\mu$m thick \pips{} detector, whereas the 500- and 1000-$\mu$m thick \pips{} detectors were paired with mini-orange spectrometers configured with five N42 \magnet{1}{1}{1/8} magnets.}
    \label{fig:ICESPICE in SE-SPS Chamber}
\end{figure}

Panel (a) of Fig.\,\ref{fig:207Bi electron-gamma coincidence matrix} shows the time-difference, $\Delta t$, spectrum between the CeBr$_3$ and \pips{} detectors, defined relative to the $\gamma$-ray detector. A coincidence peak is clearly visible with a comparably large width of $\sim$50~ns and an apparent double-hump structure. A coincidence gate was applied around this peak to suppress random background events and isolate the true $\gamma$–$e^{-}$ coincidences, see dashed lines in Fig.\,\ref{fig:207Bi electron-gamma coincidence matrix}\,(a). Energy vs. time-difference spectra are shown in Figs.\,\ref{fig:207Bi electron-gamma coincidence matrix}\,(b) and (c), respectively. The $\gamma$-$e^{-}$ matrix, when a $25 \ \mathrm{ns}\leq \Delta t \leq 75 \ \mathrm{ns}$ gate is applied, highlights several prominent decay paths of \nuc{207}{Bi} [see Fig.\,\ref{fig:207Bi electron-gamma coincidence matrix}\,(d)]. Horizontal bands correspond to conversion electrons, while vertical structures corresponds to the coincident $\gamma$ rays emitted in the same decay sequence. Due to the optimized transmission probability for $\sim 1$\,MeV electrons, the most prominent feature is the detection of electrons from the 1064-keV transition (K, L, and M electrons) in coincidence with the 569-keV $\gamma$ ray. Additional, though much weaker, coincidences are observed where the 1064-keV $\gamma$ ray is detected together with electrons from the 569-keV transition (K, L, and M electrons). A similar, but less intense, structure can be seen for the 1770-keV $\gamma$ ray in coincidence with electrons from the 569-keV transition. These transitions are shown in the level scheme in Figure~\ref{fig:207Bi_decay}. Projections of the $\gamma$-$e^{-}$ matrix are shown in panels (e) and (f). The asymmetric tail of the time-difference distribution was investigated further, see Fig.\,\ref{fig:207Bi_new_timing}. Using different gating conditions, we established that the tail of the timing distribution likely arises from electron events which do not deposit their full energy in the \pips{} detector. When, {\it e.g.}, a gate is set on the 1064-keV $\gamma$ ray, a narrow timing peak with a FWHM of $\Delta t = 5.3(4)$\,ns is recovered. This FWHM is consistent with the narrower peak observed in Fig.\,\ref{fig:207Bi electron-gamma coincidence matrix}\,(a) [see also Fig.\,\ref{fig:207Bi_new_timing}]. Based on this observation, we applied a two-dimensional gate in the $e^{-}-{\Delta}T$ matrix to test whether additional background reduction could be achieved by discarding events with partial energy deposition [see Fig.\,\ref{fig:207Bi electron-gamma coincidence matrix}\,(e) and (f)]. Indeed, some background gets reduced, with a more prominent effect in the electron spectrum.

\begin{figure}[t]
    \centering
    \includegraphics[width=0.4\linewidth]{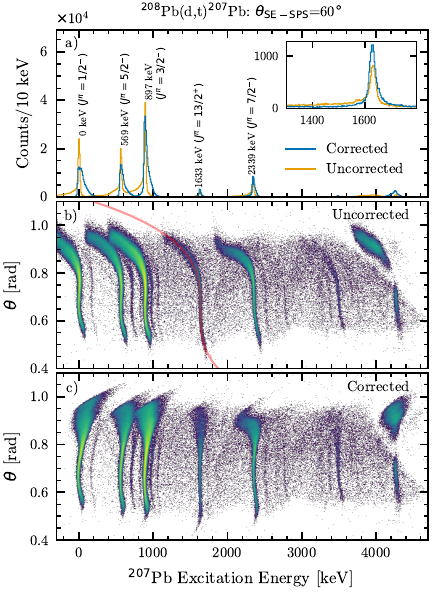}
    \caption{a) Corrected and uncorrected focal-plane (FP) spectra from the \nuc{208}{Pb}$(d,t)$\nuc{207}{Pb} reaction with the SE-SPS slits open to 12.8~msr. Operating the SE-SPS at this large of a solid angle introduces a correctable degradation in particle energy resolution due to higher-order aberrations. An inset provides a magnified view of the 1.633-MeV, $J^{\pi}=13/2^{+}$ state to illustrate the improvement after correction. b) Trajectory angle $\theta$ versus focal-plane position. A third-order polynomial was fitted to the 1.633-MeV state to correct for the aberrations. c) Corrected focal-plane spectrum after applying the $\theta$-dependent correction shown in (b).}
    \label{fig:207Pb spectrum}
\end{figure}

\section{Particle-$e^{-}$ coincidence test experiment with the ICESPICE demonstrator at the FSU SE-SPS} \label{sec: ICESPICE+SE-SPS}

\subsection{\nuc{208}{Pb}(d,$te^{-})$\nuc{207}{Pb} Experiment}

The \nuc{208}{Pb}$(d,t)$\nuc{207}{Pb} experiment was carried out at the John D. Fox Accelerator Laboratory of Florida State University using the Super-Enge Split-Pole Spectrograph (SE-SPS)~\cite{spieker_nuclear_2024}. A continuous 16-MeV deuteron beam, delivered by the 9-MV FN tandem accelerator, impinged on a 191-$\mu$g/cm$^2$ \nuc{208}{Pb} target backed with 19-$\mu$g/cm$^2$ of \nuc{12}{C}. The SE-SPS was placed at a laboratory scattering angle of 60\degree, where the cross section for populating the long-lived 1.633-MeV, $J^{\pi}=13/2^+$ state in \nuc{207}{Pb} is enhanced under the chosen kinematic conditions~\cite{muehllehner_208mathrmpb_1967}.

\begin{figure}[t]
    \centering
    \includegraphics[width=0.6\linewidth]{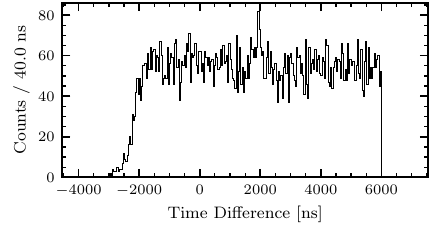}
    \caption{Time-difference spectrum between the 1000-$\mu$m thick \pips{} detector coupled to a five N42 \magnet{1}{1}{1/8} magnet mini-orange spectrometer (MOS) and the SE-SPS focal-plane scintillator. The spectrum corresponds to the first four (569-, 897-, 1633-, and 2339-keV) states populated in the \nuc{208}{Pb}$(d,t)$\nuc{207}{Pb} reaction. A distinct prompt coincidence peak is observed at 1951~ns with a FWHM of 92~ns, obtained from a Gaussian fit performed on a linear background, demonstrating successful particle-$e^{-}$ timing correlation.}
    \label{fig:207Pb_coincident_timing}
\end{figure}

ICESPICE was installed inside the SE-SPS scattering chamber for the first in-beam particle-$e^-$ coincidence test. Three mini-orange spectrometers, each coupled to a \pips{} detector of different thickness, were mounted at backward angles. All spectrometers shared the same geometry with $f = 70$ mm and $g = 25.4$ mm. The 300-$\mu$m thick detector was paired with a MOS using five N42 \magnet{1}{1}{1/16} magnets, while the 500- and 1000-$\mu$m detectors used MOS configurations with five N42 \magnet{1}{1}{1/8} magnets, optimized for $\sim$1 MeV electrons. A photograph of ICESPICE inside the SE-SPS chamber is shown in Figure~\ref{fig:ICESPICE in SE-SPS Chamber}.

Data was collected over a 50-hour run with an average beam current of 30~pnA. No external trigger was used due to the long half-life of the 1.633-MeV state, $T_{1/2} =0.806(5)$\,s~\cite{ENSDF}, and the previously unknown timing between the \pips{} detectors and the focal-plane scintillator. During the run, approximately 520,000 events were recorded for the \nuc{208}{Pb}$(d,t)$\nuc{207}{Pb} reaction. Of these, around 16,000 events correspond to the population of the 1.633-MeV state, shown in Fig.~\ref{fig:207Pb spectrum}\,(a). These events were constructed using a 3-$\mu$s coincidence window focusing on prompt coincidences with conversion electrons rather than on delayed coincidences. With the SE-SPS slits fully open, i.e., $\Delta \Omega =12.8$\,msr, higher-order aberrations degraded the energy resolution of the SE-SPS significantly. This effect is visualized in Fig.\,\ref{fig:207Pb spectrum}\,(b), where a third-order polynomial was fitted to the trajectory angle $\theta$ distribution of the 1.633-MeV peak to correct for the higher-order aberrations offline. The result of this correction is shown in panels (a) and (c) of Fig.\,\ref{fig:207Pb spectrum}. In the uncorrected data, 1.633-MeV state spans across an energy region of approximately 1130–1720~keV (a total width of $\sim$590~keV), whereas after the correction this region ranges from 1490 to 1740~keV corresponding to a significant improvement of the width to $\sim$250~keV. This is still a significant degradation of the energy resolution compared to when the SE-SPS is used with $\Delta \Omega = 4.6$\,msr \cite{spieker_nuclear_2024}, where energy resolutions of 30\,keV to 50\,keV have been achieved.

Establishing the prompt coincidence timing between ICESPICE and the SE-SPS was a critical first step. Events for ICESPICE were built between $-3000$~ns and $+6000$~ns relative to the SE-SPS focal-plane scintillator. If multiple events were detected in this window, the lowest relative time was chosen. Only events with a deposited energy of greater than 10~keV were retained to suppress noise from low-energy electrons scattering within the SE-SPS chamber into our detectors. As shown in Fig.~\ref{fig:207Pb_coincident_timing}, a weak prompt coincidence peak is observed at a time difference of $\sim$2000~ns between the 1000-$\mu$m \pips{} and the focal-plane scintillator and when gating on the first four excited (569-, 897-, 1633-, and 2339-keV) states populated in the \nuc{208}{Pb}$(d,t)$\nuc{207}{Pb} reaction. The prompt coincidence peak has a FWHM of 92~ns, which was determined from a Gaussian fit performed on a linear background.

Attempts were also made to determine the relative electron decay intensities of the excited states in \nuc{207}{Pb} by comparing the measured number of counts in the spectrometer to expectations based on the simulated $4\pi$ transmission probabilities, the known conversion coefficients \cite{ENSDF}, and the counts in the focal plane (Fig.~\ref{fig:207Pb spectrum}). For the 569-, 897-, 1633-, and 2339-keV states, almost negligible counts of approximately 1, 11, 7, and 1 electron events, respectively, were expected in the 1000-$\mu$m thick detector after the 50-hour long run. The measured spectra gated in the region of the observed prompt timing peak did not reproduce these expected yields regardless of different background subtraction methods. The observed discrepancies can be attributed to the combined effects of limited statistics from the comparably small cross sections of the \nuc{208}{Pb}$(d,t)$\nuc{207}{Pb} reaction, the reduced geometric efficiency due to only using a detector with a 50-mm$^2$ active area, and electrons that got scattered in the chamber and reached the \pips{} detectors contributing to the low-energy in-beam electron continuum. We can, however, conclude that for in-beam experiments \pips{} detectors with a larger active area and of at least a thickness of 1000 $\mu$m should be used if one plans to detect electrons with energies in excess of 1\,MeV and to acquire meaningful coincidence counts in a reasonable amount of beamtime.

\section{Summary and Outlook}

\begin{figure}[t]
    \centering
    \includegraphics[width=0.6\linewidth]{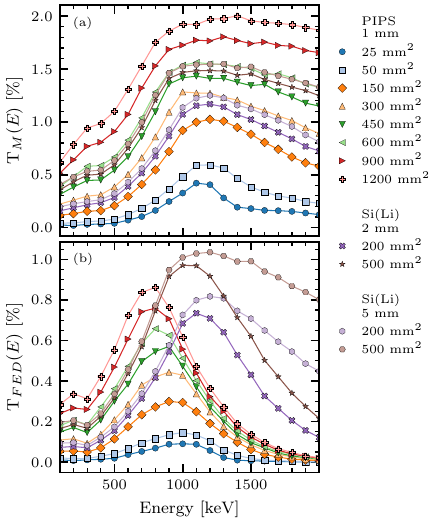}
    \caption{Geant4 simulations comparing (a) the transmission probability, $T_{M}(E)$~[\%], and (b) the full energy deposited transmission probability, $T_{\mathrm{FED}}(E)$~[\%], for a range of electron detectors. \pips{} detectors with different active areas and compatible with the current ICESPICE geometry were modeled, each with a thickness of 1~mm (1000~$\mu$m). Room-temperature Si(Li) detectors offered by Mirion \cite{noauthor_sili_nodate}, available with 200 mm$^{2}$ and 500 mm$^{2}$ active areas and with thicknesses of 2 mm and 5 mm, were also included for comparison. All simulations used the present ICESPICE mini-orange configuration consisting of five N42 magnets and the geometry with $f = 70$ mm and $g = 30$ mm. For all detectors, the manufacturer specified dead-layer thicknesses were incorporated into the geometry. A point source emitting electrons isotropically was used, and the incident energy was varied in 100 keV steps. At each energy, $10^{6}$ electrons were simulated.}
    \label{fig:detector_comparison}
\end{figure}

We have developed the ICESPICE demonstrator for internal conversion electron spectroscopy in coincidence with both $\gamma$ rays and light ions at the Super-Enge Split-Pole Spectrograph (SE-SPS) at Florida State University. ICESPICE is based on a modular mini-orange spectrometer design, employing cost-effective permanent magnets and room-temperature \pips{} detectors. Optimization of magnet configuration and geometry was carried out using SolidWorks, \comsol{}, and GEANT4 simulations to maximize magnetic transmission for $\sim$1 MeV electrons, while conforming to geometric constraints imposed by the SE-SPS internal sleeve and beamline hardware.

Calibration studies with a \nuc{207}{Bi} source demonstrated that conversion electrons could be detected with multiple spectrometer–detector combinations. Coincidence measurements with a CeBr$_3$ $\gamma$-ray detector of the CeBrA demonstrator validated the system’s ability to acquire $\gamma$–$e^-$ coincidences. The first in-beam test of ICESPICE was performed using the \nuc{208}{Pb}$(d,t)$\nuc{207}{Pb} reaction. Evidence for prompt particle–$e^-$ coincidences was observed by gating on the lower-lying excited states of \nuc{207}{Pb}, which are also populated in the electron capture of \nuc{207}{Bi}.

To study pathways to improve the performance of ICESPICE, Geant4 simulations were carried out with different detectors that could become part of the spectrometer. All simulations used the same magnetic configuration as the current demonstrator with five N42 magnets and the ICESPICE geometry with $f$ = 70 mm and $g$ = 30 mm. These studies indicate that while increasing the active area of 1000-$\mu$m thick \pips{} detectors significantly enhances the geometric efficiency over the detectors currently available at FSU, the corresponding loss in intrinsic energy resolution, as noted in the \pips{} data sheet \cite{noauthor_pips_nodate}, must also be considered. Room-temperature Si(Li) detectors \cite{noauthor_sili_nodate}, available in 200 mm$^2$ and 500 mm$^2$ active-area configurations and with thicknesses of 2 mm and 5 mm, provide substantially higher efficiency for full energy deposition of electrons with energies above 1 MeV. Although these room-temperature Si(Li) detectors do exhibit a larger intrinsic energy resolution than \pips{} devices, their use might be beneficial when used for coincidence experiments with the SE-SPS if electrons with higher energies need to be detected. The high selectivity of the SE-SPS greatly suppresses background electrons outside the region of interest. Therefore, the superior efficiency (stopping power) of thick room-temperature Si(Li) detectors could outweigh their poorer resolution, making them also an option for high energy internal conversion electron measurements with ICESPICE. Future work will explore different magnet strengths, magnet layouts, and variations of the $f$ and $g$ distances to improve the transmission for each detector type and for electron energies much smaller than 1\,MeV as often encountered in the lanthanides and actinides. Future upgrades will also consider integrating room-temperature Si(Li) detectors into the ICESPICE system in order to improve sensitivity to high energy conversion electrons.

Additional calibration measurements with sources such as \nuc{133}{Ba} or \nuc{152}{Eu} will provide further benchmarks at lower energies. Direct measurements of the magnetic field would also allow for a more accurate mapping in \comsol{} and an even more improved reproduction of the transmission probability in the GEANT4 simulations. Additional background suppression will also be explored by biasing the SE-SPS beam dump and applying shielding materials to reduce the flux of low-energy electrons reaching the detector face from scattering within the SE-SPS chamber.

\section*{Acknowledgments}
This material is based upon work supported by the U.S. National Science Foundation (NSF) under Grant Nos. PHY-2012522 (FSU) [WoU-MMA: Studies of Nuclear Structure and Nuclear Astrophysics] and PHY-2412808 (FSU) [Studies of Nuclear Structure and Nuclear Astrophysics], as well as by the Department of Energy, National Nuclear Security Administration, under Award No. DE-NA0004150 through the Center for Excellence in Nuclear Training And University-based Research (CENTAUR). Additional support by Florida State University is gratefully acknowledged. This research used targets provided by the Center for Accelerator Target Science at Argonne National Laboratory, a DOE Office of Science User Facility supported by the U.S. Department of Energy, Office of Nuclear Physics, under Award No. DE-AC02-06CH11357. We also gratefully acknowledge J. Aragon, P. Barber, R. Boisseau, and B. Schmidt for their assistance.

\bibliographystyle{apsrev4-1}
\bibliography{ICESPICE_NIM_A}

\end{document}